\documentclass[a4paper,fleqn,usenatbib]{mnras}

\usepackage{newtxtext,newtxmath}

\usepackage[T1]{fontenc}
\usepackage{ae,aecompl}

\usepackage{graphicx}	
\usepackage{amsmath}	

\usepackage[dvipsnames]{xcolor} 

\newcommand{\cob}{\texorpdfstring{CO$^5$BOLD}{CO5BOLD}}
\newcommand{\sta}{Stagger}

\title[The power spectrum of solar granulation]{The signature of granulation in a solar power spectrum as seen with \cob} 

\author[M. S. Lundkvist et al.]{
Mia S. Lundkvist$^{1,2}$\thanks{E-mail: lundkvist@phys.au.dk},
Hans-G{\"u}nter Ludwig$^{2}$,
Remo Collet$^{1}$
and Thomas Straus$^{3}$
\\
$^{1}$Stellar Astrophysics Centre, Aarhus University, Ny Munkegade 120, 8000 Aarhus C, Denmark\\
$^{2}$Zentrum f{\"u}r Astronomie der Universit{\"a}t Heidelberg, Landessternwarte, K{\"o}nigstuhl 12, 69117 Heidelberg, Germany\\
$^{3}$INAF, Osservatorio Astronomico di Capodimonte, Via Moiarello 16, 80131 Napoli, Italy
}

\date{Accepted 2020 November 18. Received 2020 November 18; in original form 2020 August 28}

\pubyear{2020}

\begin{document}
\label{firstpage}
\pagerange{\pageref{firstpage}--\pageref{lastpage}}
\maketitle

\begin{abstract}
The granulation background seen in the power spectrum of a solar-like oscillator poses a serious challenge for extracting precise and detailed information about the stellar oscillations. Using a 3D hydrodynamical simulation of the Sun computed with \cob, we investigate various background models to infer, using a Bayesian methodology, which one provides the best fit to the background in the simulated power spectrum.
We find that the best fit is provided by an expression including the overall power level and two characteristic frequencies, one with an exponent of 2 and one with a free exponent taking on a value around 6.
We assess the impact of the 3D hydro-code on this result by repeating the analysis with a simulation from Stagger and find that the main conclusion is unchanged. However, the details of the resulting best fits differ slightly between the two codes, but we explain this difference by studying the effect of the spatial resolution and the duration of the simulation on the fit. Additionally, we look into the impact of adding white noise to the simulated time series as a simple way to mimic a real star. We find that, as long as the noise level is not too low, the results are consistent with the no-noise case.
\end{abstract}




\section{Introduction}
When fitting the power spectrum of a solar-like oscillator in order to extract information about the oscillations, it is important to properly account for the stellar background attributable to near-surface convection, since this may otherwise bias the results \citep[see, for instance,][]{ref:kallinger2014,ref:kamiaka2019}. Several different descriptions of the background exist with slightly varying components, but common to all of them is a contribution from white noise in addition to the stellar background, which consists of a contribution from stellar activity as well as at least one component arising from surface granulation.

A classic background model was proposed by \citet{ref:harvey1985} based on the flux coming from a granule being described as an instantaneous rise followed by an exponential decay, resulting in a Lorentzian profile in the power spectrum. He suggested that the background be composed of three granulation components in addition to a component from activity and one from white noise. This description was later modified to allow for a variable exponent $c_i$ \citep{ref:harvey1993}:
\begin{equation}
	\label{eq:harvey}
	B(\nu) = \sum\limits_{i=0}^{4} \frac{A_i}{1 + \left( 2 \pi \nu \tau_i \right)^{c_i}} + B_0 \ .
\end{equation}
Here, $A_i$ is the power, $\nu$ is the frequency, $\tau_i$ is the characteristic time scale and $B=0$ the flat white noise contribution. Modifications to eq.~\ref{eq:harvey} have been proposed by, for instance, \citet{ref:karoff2008} and tested by, for example, \citet{ref:mathur2011} and \citet{ref:kallinger2014}.

\citet{ref:kallinger2014} used \textit{Kepler} data of red giants to compare eight different background models and determine which was the best fit to the data using bayesian model comparison. They favoured a model with two background components, which was also found to be the case for main sequence solar-like stars by \citet{ref:karoff2013}. However, the true shape of the stellar background remains elusive due to the fact that convection and its imprint on the power spectrum is not fully understood \citep[see, for instance,][and references therein]{ref:kallinger2014}

In order to tackle this question from a different angle, we have investigated how well nine background models fit a simulated solar power spectrum made with the 3D hydrodynamics code \cob\ (COnservative COde for the COmputation of COmpressible COnvection in a BOx of L Dimensions, L=2,3) \citep[][used here with L=3]{ref:freytag2012}. The \cob\ solar simulation is described in section~\ref{sec:method} along with the background models and the fitting that we use. The results of the model comparison is the topic of section~\ref{sec:results}, while we discuss the results and the tests that we have carried out to ensure that our method is robust in section~\ref{sec:discussion}, and we finish with our conclusions in section~\ref{sec:conclusion}.

\section{Method}
\label{sec:method}

\subsection{\cob\ solar simulation}
\label{subsec:simulations}
\begin{figure}
	\includegraphics[width=\columnwidth]{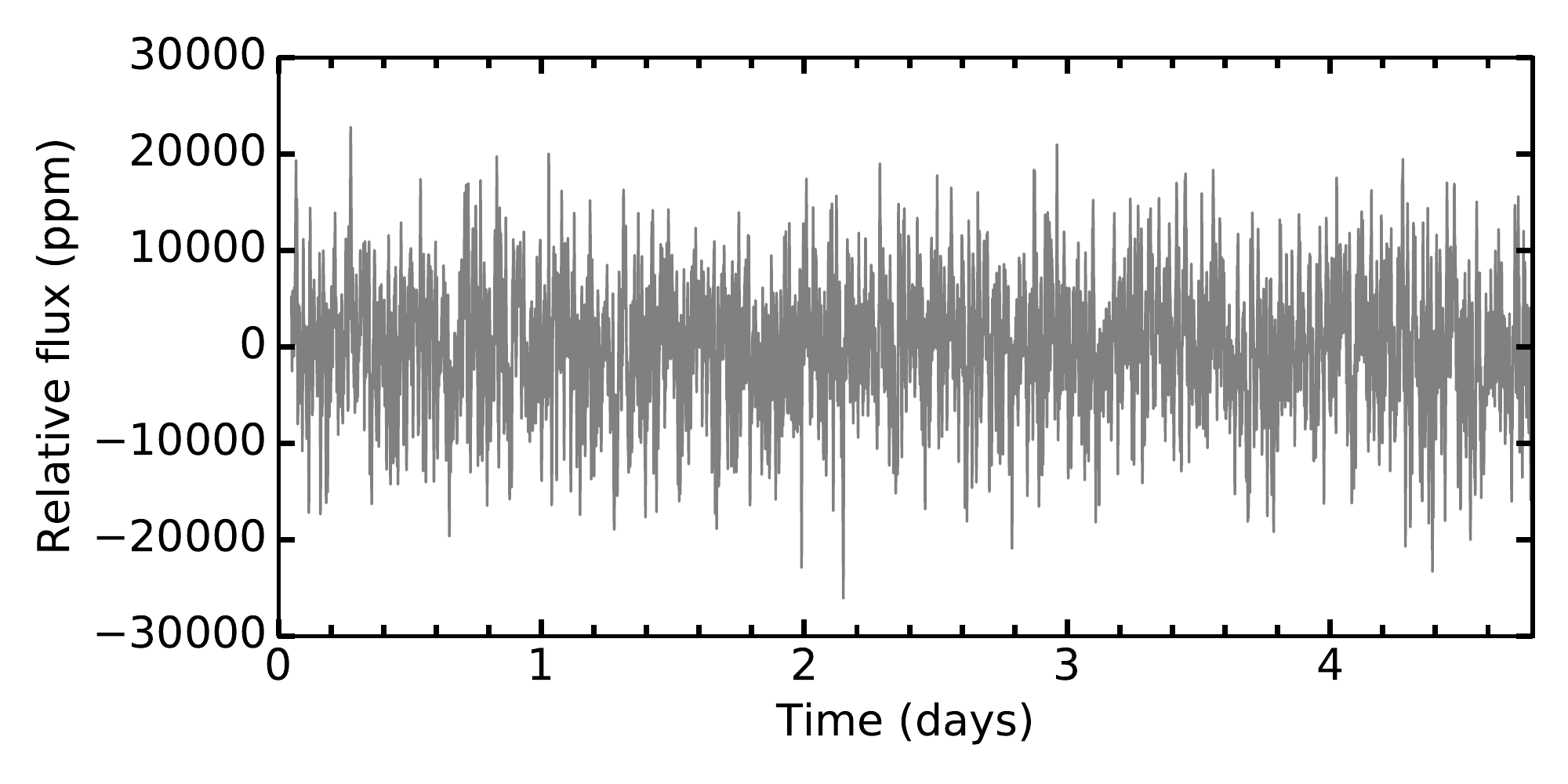}\\
	\includegraphics[width=\columnwidth]{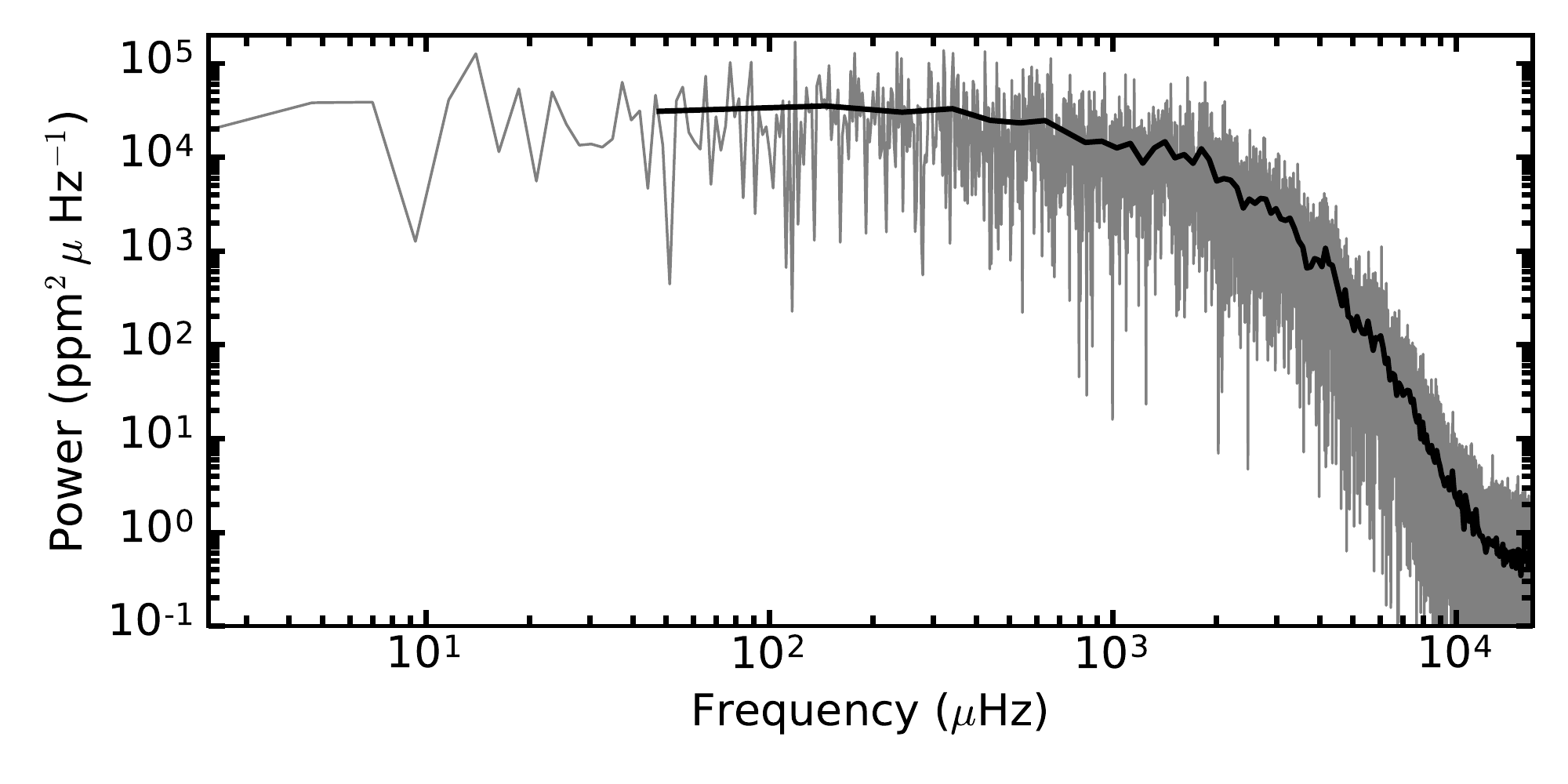}
    \caption{\textit{Top:} \cob\ simulated time series of the Sun (the Reference model). \textit{Bottom:} Power spectrum computed from the \cob\ time series with a heavily smoothed version overplotted in black.}
    \label{fig:co5boldts}
\end{figure}
We have used a solar simulation from \cob\ with (mostly) 30 second sampling lasting for a total of ${\sim}4.5$ days of solar time. 
The radiation-hydrodynamics code \cob\ solves the equations of hydrodynamics coupled to the radiative transfer equation for a stratified, chemically homogeneous medium. In the present context, we are using the 'box-in-a-star' set-up where a small but representative volume located in the solar surface layers is simulated. The bottom boundary allows for the free in- and outflow of matter, the top boundary is transparent to wave motions, while the lateral boundaries are periodic. All boundaries are transparent to radiation. The total flux permeating the simulation volume is controlled by the heat content of material entering the box from below. \cob\ stores periodically the full flow state, and so-called mean-files, which contain horizontal averages of various flow and radiation quantities. The mean-files are used here for further analysis. More information on the code can be found in \citet{ref:freytag2012}.

From the \cob\ mean-files we extracted the model time (in seconds) and the horizontal mean of the emergent flux in the simulation (in erg cm$^{-2}$ s$^{-1}$). We then turned this into relative flux in ppm (parts per million) similar to what is often done with observations \citep[e.g. from \textit{Kepler}, see, for instance, eq.~6 of][]{ref:handberg2014}.

Using this flux time series, we have computed a power density spectrum using the fast implementation of the Lomb-Scargle periodogram \citep{ref:lomb1976,ref:scargle1989} by \citet{ref:press1989}. Note that we have chosen to study a single patch of the Sun and thus not create a disk-integrated power spectrum as detailed by \citet{ref:ludwig2006} as this in our case does not impact the overall shape of spectrum, which is what we are interested in (see appendix~\ref{app:shape} for further details). The flux time series and the power spectrum are shown in Fig.~\ref{fig:co5boldts}. In the following, we use this Reference model power spectrum to test which of nine different background models (see Table~\ref{tab:models}) that best reproduce the shape of the power spectrum.

\subsection{Background models}
\label{subsec:backgroundmodels}
The set of background models, $\mathcal{B}$, that we are testing are from \citet{ref:kallinger2014} with the addition of a single model (model I). The nine models consist of five models with one granulation component and four with two components, and the first four and the following four have the same functional form of the granulation component(s). Table~\ref{tab:models} lists all the background models used here.
\begin{table}
	\centering
	\caption{The background models, $\mathcal{B}_i(\nu), i \in \lbrace A, B, \dots , I \rbrace$, use in this text. Note that \citet{ref:kallinger2014} include a $\xi$ also on models D and H.}
	\label{tab:models}
	\begin{tabular}{lcr} 
		\hline
		Model & Equation & Reference\\
		\hline
		A & $\frac{\xi a^2/b}{1 + \left( \nu/b \right)^2}$ & \citet{ref:kallinger2014}\\
		B & $\frac{\xi a^2/b}{1 + \left( \nu/b \right)^4}$ & \citet{ref:kallinger2014}\\
		C & $\frac{\xi a^2/b}{\left[ 1 + \left( \nu/b \right)^2 \right]^2}$ & \citet{ref:kallinger2014}\\
		D & $\frac{a^2/b}{1 + \left( \nu/b \right)^l}$ & \citet{ref:kallinger2014}\\
		E & $\frac{\xi a^2/b}{1 + \left( \nu/b \right)^2} +
		     \frac{\xi c^2/d}{1 + \left( \nu/d \right)^2}$ & \citet{ref:kallinger2014}\\
		F & $\frac{\xi a^2/b}{1 + \left( \nu/b \right)^4} +
		     \frac{\xi c^2/d}{1 + \left( \nu/d \right)^4}$ & \citet{ref:kallinger2014}\\
		G & $\frac{\xi a^2/b}{\left[ 1 + \left( \nu/b \right)^2 \right]^2} +
		     \frac{\xi c^2/d}{\left[ 1 + \left( \nu/d \right)^2 \right]^2}$ & \citet{ref:kallinger2014}\\
		H & $\frac{a^2/b}{1 + \left( \nu/b \right)^l} +
		     \frac{c^2/d}{1 + \left( \nu/d \right)^k}$ & \citet{ref:kallinger2014}\\
		I & $\frac{a^2}{1 + \left( \nu/b \right)^2 + \left( \nu/d \right)^k}$ & This work\\
		\hline
	\end{tabular}
\end{table}
For each background model, $a$ and $c$ are the amplitudes of the granulation components, $b$ and $d$ designate the characteristic frequencies, $\nu$ is the frequency, $l$ and $k$ are exponents and $\xi$ is a normalisation factor \citep[chosen as in][]{ref:kallinger2014} such that the power of the granulation component corresponds to the area under the super-Lorentzian function in the power density spectrum: $\int\limits_0^\infty (\xi/b)/(1 + (\nu/b)^c)d\nu = 1$. For models D, H and I, the value of $\xi$ can not be determined analytically, and thus we simply omitted the $\xi$ factor as we are not concerned with the actual amplitude values in the current work.

It can be seen that models A and E consist of the classical Harvey model, models B and F have super-Lorentzian component(s) with the exponent fixed to $4$, models C and G have a slightly different functional form and models D and H also consist of super-Lorentzian component(s), but with a free exponent. Model I is a hybrid between the single and two-component models in that it only has a single component, but contains two characteristic frequencies with each their associated drop-off slope (2 and free respectively).

Since we are working with a simulation and not actual observations, we have changed a few of the details with respect to the background fit as compared to \citet{ref:kallinger2014}. First, we do not incorporate an apodization factor \citep[due to a finite integration time,][]{ref:chaplin2011}, since our time series points are not integrated over the sampling time, but 'instantaneous' measurements. Second, we do not include an activity term in our background description, since the simulation does not include this feature. Third, no Gaussian is included to take the solar-like oscillations into account as these are not significantly excited in our power spectrum (we have a box modes of low height that we disregard). Fourth, we do not include a white-noise component in our background model as our simulation does not contain this either \citep[see, for instance][]{ref:frandsen2007}. Instead we include the mirroring effect around the Nyquist frequency, $\nu_\mathrm{Nyq}$, since this, due to the absence of white noise, is clearly seen in the simulated power spectrum. As a consequence, what is actually fitted to the power spectrum is of the following form:
\begin{equation}
	\label{eq:totback}
	B_i(\nu) = \mathcal{B}_i(\nu) + \mathcal{B}_i(2 \cdot \nu_\mathrm{Nyq} - \nu), 
													\qquad i \in \lbrace A, B, \dots , I \rbrace ,
\end{equation}
with the background models $\mathcal{B}$ listed in Table~\ref{tab:models}.

\subsection{Fitting}
\label{subsec:fitting}

We fit the background models in Table~\ref{tab:models} (through eq.~\ref{eq:totback}) to our $10$-point binned power spectrum in a Bayesian manner similarly to what was done by \citet{ref:davies2016}. In order to do this and be able to obtain estimates of the parameters, we need, in addition to the model, priors, a method for evaluating the likelihood and an MCMC sampler that will allow us to sample the posterior probability density ($p(\mathbf{\Theta} \vert D, M)$). The posterior probability density for a set of parameters ($\mathbf{\Theta}$), given observed data ($D$) and a model ($M$) is given by Bayes theorem:
\begin{equation}
	\label{eq:bayestheo}
	p(\mathbf{\Theta} \vert D, M) = \frac{
		p(\mathbf{\Theta} \vert M) p(D \vert \mathbf{\Theta}, M)		
		}
		{p(D \vert M)} \ .
\end{equation}
Here, $p(\mathbf{\Theta} \vert M)$ are the prior probabilities of the model parameters, $p(D \vert \mathbf{\Theta}, M)$ is the likelihood of the data given a specific choice of parameters and $p(D \vert M)$ is the normalisation factor known as the model evidence or the marginal likelihood. It be determined by integrating the numerator in eq.~\ref{eq:bayestheo} \citep[see e.g.][]{ref:handberg2011,ref:gronau2017} over the entire parameter space:
\begin{equation}
	\label{eq:evidence}
	p(D \vert M) = \int\limits_\mathbf{\Theta}
						p(\mathbf{\Theta} \vert M) p(D \vert \mathbf{\Theta}, M)
						d\Theta \ .
\end{equation}

From the posterior probability density, parameter estimates can be obtained by marginalising over (i.e. integrating out) the remaining parameters. We take as our parameter estimate the median of the marginalised posterior probability density and use the 16\% and 84\% quantiles to give the 1-sigma uncertainties.

Below we will address our choice of priors, the MCMC sampler used, the likelihood function and the employed method for model comparison.

\subsubsection{Priors} \label{subsubsec:priors}

Very similar and only weakly informative priors were used for each of the parameters in the background models (see appendix~\ref{app:priors} for details). For the amplitudes and characteristic frequencies ($a$, $b$, $c$ and $d$) these were log-normal priors, while for the exponents ($l$ and $k$) they were normal distributions. The only parameter for which we used an informative prior was the Nyquist frequency, where we opted for a normal distribution with a width of $0.001 \cdot \nu_\mathrm{Nyq} \approx 16.4 \ \muup$Hz. This was done because we know the value of the Nyquist frequency well for our simulation, and we found that if we allowed it to vary too much, some of the chains would land on an un-physical value. This was caused by the fact that by changing the Nyquist frequency, because of the mirroring effect around this point in the power spectrum, the slope of the high-frequency end of the model power spectrum could be changed, which would allow for a better fit for a given background model with a predefined exponent.

\subsubsection{MCMC details} \label{subsubsec:mcmc}

For the MCMC sampling, we use the Python interface to Stan \textit{PyStan}\footnote{\url{https://pystan.readthedocs.io/en/latest/}}, which employs the No-U-Turn sampler variant of Hamiltonian Monte Carlo. This provides the posterior distributions for the parameters. We employ 4 chains each taking $50,000$ steps, half of which is taken as the burn-in and subsequently removed. After the burn-in has been removed, each chain trace is inspected to make sure that a sufficient part of the chain has been removed. It was also checked if the posterior distributions showed a close resemblance to the priors, something which happens when the data are not informative.

\subsubsection{The likelihood function} \label{subsubsec:likelihood}

For numerical stability, we map the log of the likelihood using the expression by \citet{ref:handberg2011} for uncorrelated frequency bins ($j$) in the power spectrum:
\begin{equation}
	\label{eq:loglike}
	\ln p(D \vert \mathbf{\Theta}, M) = \sum\limits_j \ln \left( f(D_j, \mathbf{\Theta}, M_j) \right) \ .
\end{equation}
Given that we have binned our power spectrum, it will obey a $\chi^2$ probability distribution with $2s$ degrees of freedom with $s$ being the number of binned data points. Thereby, we have the probability density \citep{ref:appourchaux2003,ref:appourchaux2004,ref:handberg2011}:
\begin{equation}
	\label{eq:chi2pdf}
	f(D_j, \mathbf{\Theta}, M_j) =
						\frac{s^{s-1}}{\left(s-1\right)!}
						\frac{D_j^{s-1}}{M_j\left(\mathbf{\Theta}\right)^s}
						\exp \left(- \frac{sD_j}{M_j\left(\mathbf{\Theta}\right)} \right) \ .
\end{equation}
This expression can then be combined with eq.~\ref{eq:loglike} to yield the log-likelihood. 

\subsubsection{Model comparison} \label{subsubsec:comparison}

To compare models in a Bayesian manner, it is crucial to know the marginal likelihood, i.e., the normalising constant in the posterior probability density (eq.~\ref{eq:bayestheo}).

The model probability for model $\mathcal{M}_i$, $i \in \lbrace 1, 2, \dots ,m \rbrace$, given the data $D$ can be obtained as \citep{ref:berger2005,ref:gronau2017b}
\begin{equation}
	\label{eq:modelprob}
	p\left( \mathcal{M}_i \vert D\right) = \frac{
			p\left( D \vert \mathcal{M}_i\right) p \left(\mathbf{\Theta} \vert \mathcal{M}_i \right)
			}{
			\sum\limits_{j=1}^m p\left( D \vert \mathcal{M}_j \right)
			p\left(\mathbf{\Theta} \vert \mathcal{M}_j \right) } \ .
\end{equation}
Here, $p\left(\mathbf{\Theta} \vert \mathcal{M}_i \right)$ can be recognised as the prior model probability, and $p\left( \mathbf{D} \vert \mathcal{M}_i\right)$ as the marginal likelihood (model evidence) of model $\mathcal{M}_i$. When comparing models, eq.~\ref{eq:modelprob} allows us to obtain the relative plausibility of a given model, i.e. the change in beliefs regarding the relative plausibility of the models induced by the data.

Assuming that our priors will not guide us to prefer one model over another, we can disregard the priors in eq.~\ref{eq:modelprob} \citep{ref:handberg2011} whereby we end up with $p_i \equiv p\left( \mathcal{M}_i \vert D\right) = z_i / \sum\limits_j z_j$ with $z$ being the marginal likelihood normalised by a reference value $z_0$ to simplify the computations \citep[following][]{ref:kallinger2014}. Thus, in order to be able to compare the competing background models we compute the log marginal likelihood using our own python implementation of the \textit{bridgesampling} package\footnote{\url{https://CRAN.R-project.org/package=bridgesampling}} \citep{ref:gronau2017}.

\section{Results}
\label{sec:results}

\subsection{Fitting the \cob\ Reference model}
\label{subsec:cobresults}
\begin{figure}
	\includegraphics[width=\columnwidth]{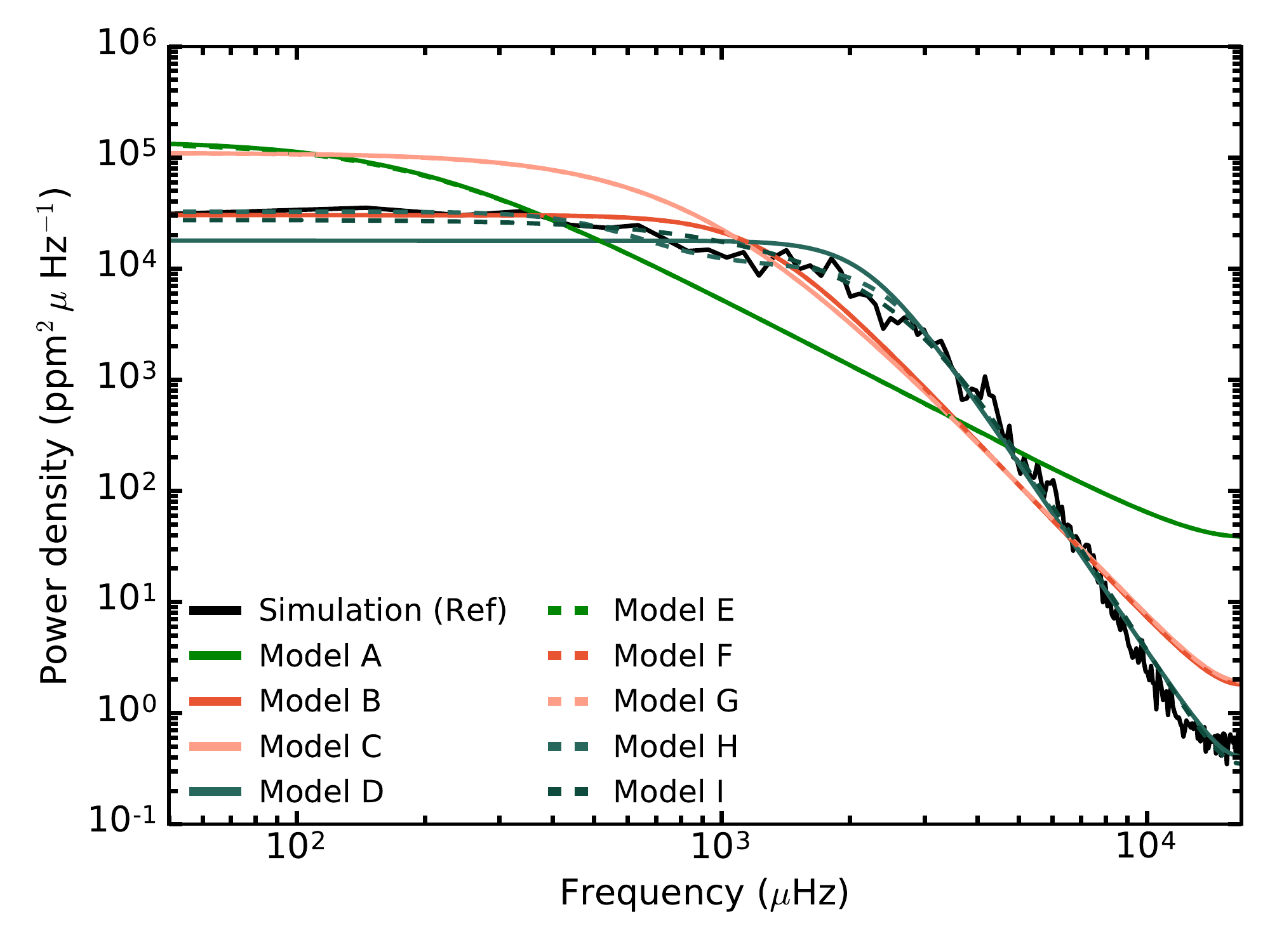}
    \caption{Smoothed simulated solar power spectrum from \cob\ (Reference) with each of the fitted background models overplotted using the median value for each parameter.}
    \label{fig:modelfit}
\end{figure}
Figure~\ref{fig:modelfit} shows each of the fitted background models plotted on top of a smoothed version of the simulated solar power spectrum. In several cases, the one- and corresponding two-component models essentially overlap, because one of the components in the two-component model has a very small amplitude. The same can be seen in the case of model B/F in \citet{ref:kallinger2014}. In fact, in the cases of models E, F and G the data do not support the addition of an extra component. This can be seen from the returned posteriors for these models, where one of the components in each case almost reproduce the input priors.

We take as our reference model, model I, which we find to be the background model that best describe our simulated solar power spectrum. This can be seen from the model probabilities listed in Table~\ref{tab:results}, which have been determined as described in Sect.~\ref{subsubsec:comparison}.
\begin{table}
	\centering
	\caption{Marginal likelihood (z) and model probability (p) for the nine different tested background models using $\ln (z_0) = -4426$ as a reference value.}
	\label{tab:results}
	\begin{tabular}{lrr} 
		\hline
		Model & $\ln \left(z/z_0\right)$ & p\\ 
		\hline
		A & $ -12990 $ & $<10^{-200}$ \\ 
		B & $ -2702 $  & $<10^{-200}$ \\ 
		C & $ -3118 $  & $<10^{-200}$ \\ 
		D & $ -255 $   & ${\sim}10^{-111}$ \\
		E & $ -12996 $ & $<10^{-200}$ \\ 
		F & $ -2701 $  & $<10^{-200}$ \\ 
		G & $ -3118 $  & $<10^{-200}$ \\ 
		H & $ -199 $   & ${\sim}10^{-86}$ \\ 
		I & $ 0.17 $   & ${\sim}1$ \\ 
		\hline
	\end{tabular}
\end{table}

\begin{figure*}
	\includegraphics[width=.8\textwidth]{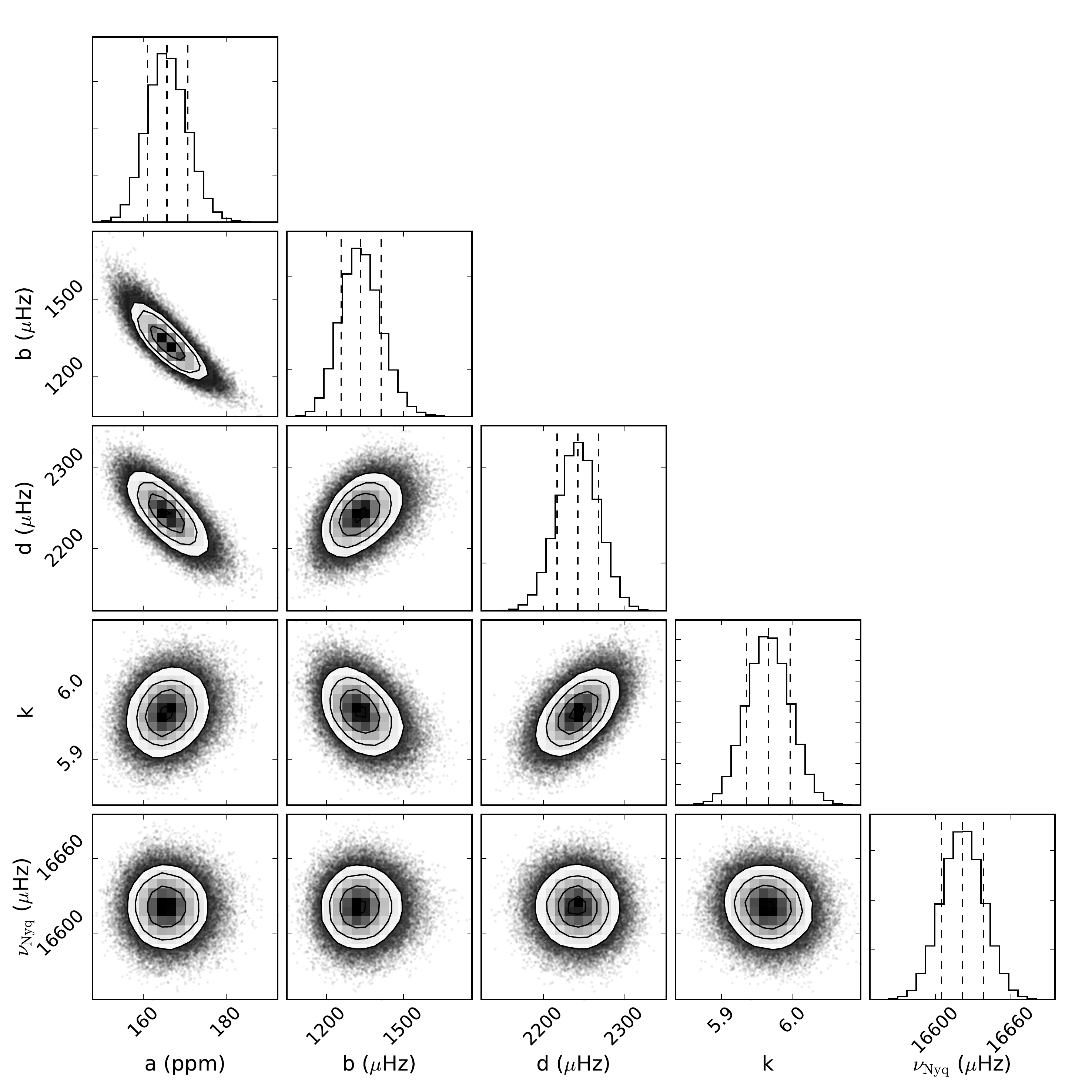}
    \caption{Corner plot for model I fit to the Reference \cob\ power spectrum showing the sampled space and contour lines (corresponding to 0.5, 1, 1.5 and 2 sigma) in the 2D plot and the median and $1\sigma$ values in the histograms. The plot shows that the amplitude and characteristic frequencies are anti-correlated.}
    \label{fig:modI_corner}
\end{figure*}

Figure~\ref{fig:modI_corner} shows a corner plot for the preferred background model I made using \textit{corner.py}\footnote{\url{https://corner.readthedocs.io/en/latest/}} \citep{ref:foremanmackey2016}. This plot shows the sampled parameter space and thus reveals any degeneracies between the parameters as well as if a parameter shows a bimodal nature. It can be seen that each of the distributions appear single-peaked with a clearly defined central value. Some correlations are present between the various parameters, in particular the overall amplitude of the granulation signal is anti-correlated with the values of the characteristic frequencies. This is not surprising given that the fitting tries to preserve total power; a shift of the cut-off to higher frequencies will then mean that the amplitude at low frequencies must become smaller. The parameter estimates and associated uncertainties of the characteristic frequencies and the free exponent are listed in the first row of Table~\ref{tab:fitparams}. The result implies that, at least for the Sun, model~I with a free exponent close to six provides the best fit to the granulation background in the intensity power spectrum.

\begin{figure}
	\includegraphics[width=\columnwidth]{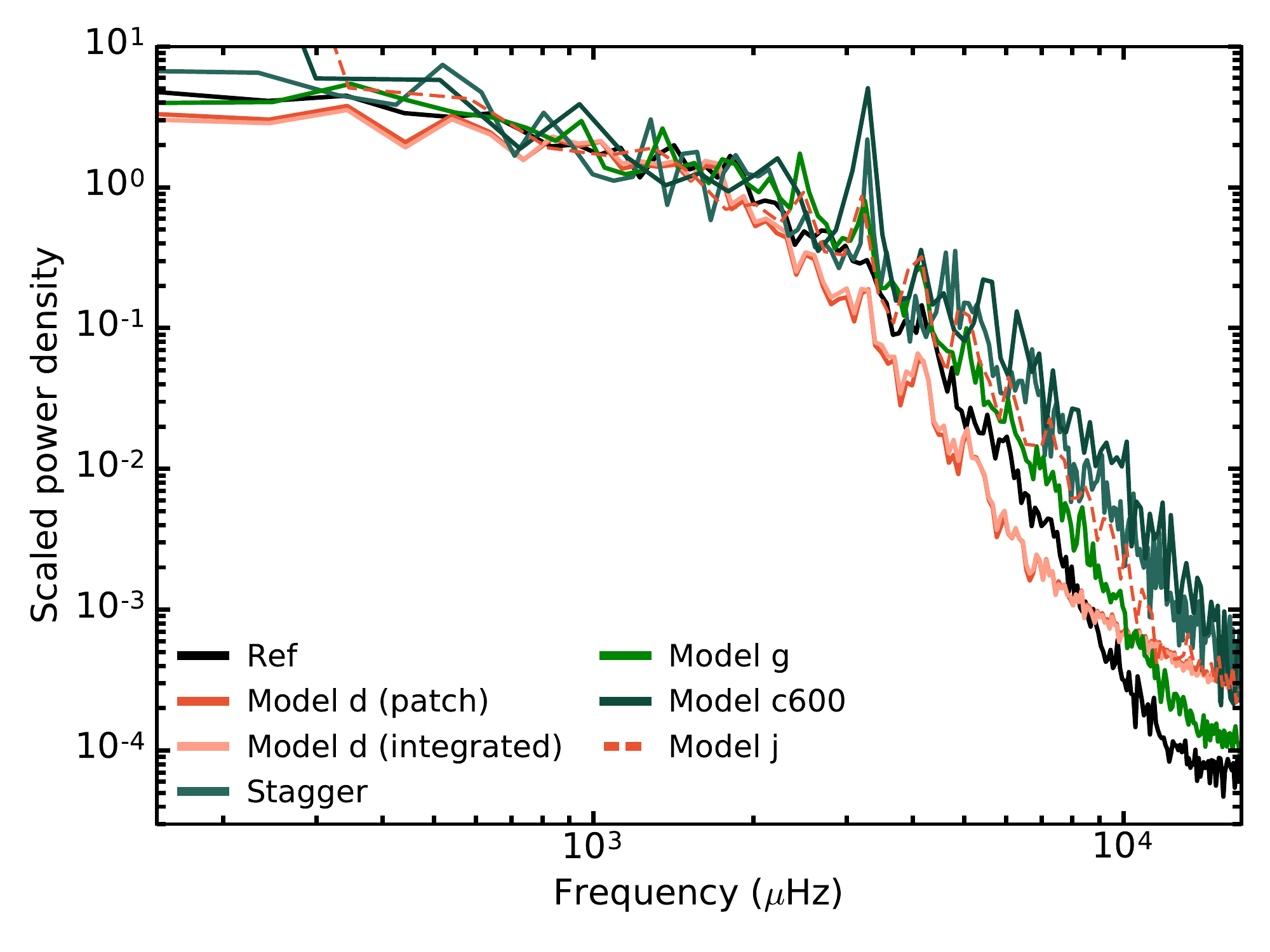}
    \caption{Smoothed version of the power spectra used in this work. They have been scaled to match in the $1-2$ mHz range. The legend gives the meaning of the different colours (refer to Table~\ref{tab:simulations}). Notice that all power spectra have been binned heavier than what was done in the fitting (and the ones from models c600 and j more than the others because of the larger amount of scatter).}
    \label{fig:comparingpower}
\end{figure}

However, it is unclear how this result is affected by the exact simulation that we have used. Therefore, in the following subsections we will describe several tests that we have carried out to assess the sensitivity of our results to the chosen simulation, in addition to investigating one key difference between real and simulated power spectra, namely the absence of white noise in the simulation. Table~\ref{tab:simulations} below gives some relevant details about the various simulations used for the tests. The corresponding power spectra can be seen in Fig.~\ref{fig:comparingpower}. Note that two power spectra are present for model d in this figure, one called 'patch' and one called 'integrated'. These two have been computed in slightly different ways (as mentioned in Sect.~\ref{subsec:simulations}) and are the subject of appendix~\ref{app:shape}. In the following, we choose to use the integrated one (similarly for the power spectra of hydro models g, j and c600), but this has no effect on the results.
\begin{table*}
	\centering
	\caption{Details about the solar simulations used in the present section. The sampling time given is the median in the case of the Reference model. The last dimension relates to the vertical direction while the first two represent the horizontal directions. The spatial resolution has been computed as $\mathrm{size}_i/N_i$ for the two horizontal directions. In the vertical direction, for all models but the Reference one, the grid has non-equidistant spacing, and the listed resolution is the one at optical depth unity (for \sta, at the optical surface). The top- and bottom five layers in the vertical direction that \sta\ uses for computing derivatives and interpolating near the boundaries have been disregarded. Also given is how the radiative transfer was treated, where '12 bin' refers to opacity binning with 12 bins.}
	\label{tab:simulations}
	\begin{tabular}{lrrrrrrr} 
		\hline
		Simulation (code) & Duration & Sampling & Nr of grid points & Size & Spatial resolution & Radiative & Sect.\\
		 & (ks) & (s) & $N_x \times N_y \times N_z$ & ($\mathrm{Mm}^3$) & ($\mathrm{km}^3$) & transfer & \\
		\hline
		Ref (\cob) & $407.9$ & 30 & $200 \times 200 \times 250$
		           & $11.2 \times 11.2 \times 5.25$
				   & $56.0 \times 56.0 \times 21.0$ & grey
				   & Reference \\
		Model d (\cob) & $264.0$ & 10 & $189 \times 189 \times 150$
		               & $18.6 \times 18.6 \times 8.5$
					   & $98.4 \times 98.4 \times 18.2$ & grey 
					   & \ref{subsec:resolution}, \ref{app:shape} \\
		Model g (\cob) & $300.0$ & 10 & $378 \times 378 \times 300$
		               & $18.6 \times 18.6 \times 8.5$
					   & $49.2 \times 49.2 \times 9.0$ & grey
					   & \ref{subsec:resolution} \\
		Model j (\cob) & $42.0$ & 10 & $534 \times 534 \times 424$
					   & $18.6 \times 18.6 \times 8.4$
					   & $34.8 \times 34.8 \times 6.4$ & grey
					   & \ref{subsec:resolution} \\
		Model c600 (\cob) & $23.4$ & 10 & $250 \times 250 \times 207$
		                  & $8.0 \times 8.0 \times 2.3$
						  & $32.0 \times 32.0 \times 10.0$ & 12 bin
						  & \ref{subsec:resolution} \\
		Stagger (\sta) & $94.4$ & 30 & $240 \times 240 \times 230$
					   & $6.0 \times 6.0 \times 3.6$
					   & $25.1 \times 25.1 \times 7.0$ & 12 bin
					   & \ref{subsec:stagger} \\
		\hline
	\end{tabular}
\end{table*}

\subsection{Comparing to \sta\ simulation}
\label{subsec:stagger}

The primary test that we have carried out, pertains to the choice of 3D hydrodynamical code. In order to assess the impact of this, we have fitted all of the background models listed in Table~\ref{tab:models} to a power spectrum made from a simulated solar time series from the \sta\ code \citep{ref:collet2011}. The Stagger hydrodynamics code uses a 'box-in-a-star' set-up with open boundaries in the vertical direction and horizontally periodic boundary conditions. In the vertical direction the grid has non-equidistant spacing with higher resolution at the photosphere. Details on the specific version used here can be found in \citet{ref:nordlund1994,ref:amarsi2018} and \citet{ref:collet2018}.
This simulated solar time series is somewhat shorter, but has a sampling that is almost identical to the one in the Reference \cob\ time series.

It is evident from Fig.~\ref{fig:comparingpower} that the box modes stand out more clearly in the \sta\ simulation than in the Reference \cob\ simulation. Because they tend to drive the background fit, we tried several different fits to investigate the best way to address their presence in the \sta\ power spectrum. We tried simply leaving them in, cutting out the region with the most prominent box modes ($2800-5800 \ \muup$Hz), and adding one or two Lorentzian profiles to account for the most prominent box mode(s). In the end, owing to the importance of the Lorentzian wings even far from the box mode(s), we decided to use the Lorentzian profile(s). We fitted all nine models, both including one and two Lorentzian components to the \sta\ power spectrum, and we found that in both cases background model~I was the preferred model, i.e. the background model yielding the highest model probability, as was the case for our Reference \cob\ simulation (and furthermore it provided the most robust fit). Table~\ref{tab:fitparams} (top two entries) lists the values of the characteristic frequencies and the free exponent determined in the fitting of the \cob\ and \sta\ power spectra. In the latter case, the parameters for model I including two Lorentzian components are given as this version presented a higher marginal likelihood than including just one Lorentzian.

\begin{table}
	\centering
	\caption{Best-fitting (median) parameters with $1\sigma$ uncertainties for the background model~I fit to the Reference \cob\ power spectrum as well as the \sta\ power spectrum and the power spectra used in Sect.~\ref{subsec:resolution}. The number in parentheses following the simulation name indicates the number of Lorentzians included in the fit to account for box modes.}
	\label{tab:fitparams}
	\begin{tabular}{lccc} 
		\hline
		Simulation (fit type) & $b$ & $d$ & $k$\\ 
		\hline
		Ref (0L)	 & $ 1332 \substack{+82 \\ -75} $ & $ 2243 \pm 26 $ & $ 5.97 \pm 0.03 $ \\
		Stagger (2L) & $ 740 \substack{+76 \\ -73} $ & $ 2919 \substack{+97 \\ -99} $ & $ 6.01 \substack{+0.09 \\ 																							-0.08} $ \\
		Model d (1L) & $ 1176 \substack{+109 \\ -94} $ & $ 1958 \substack{+37 \\ -38} $ & $ 5.84 \pm 0.08 $ \\
		Model g (0L) & $ 1471 \substack{+105\\-95} $ & $ 2505 \pm 35 $ & $ 5.84 \pm 0.04 $ \\
		\hline
	\end{tabular}
\end{table}

It can be seen from these results, that the free exponent turns out identical for the two simulations, while the characteristic frequencies are somewhat different. This means that the high frequency slope is similar in the two power spectra, but the detailed distribution of power is different. It is reassuring that although the input physics differ between the two codes, the slopes are consistent as this relates to characteristics of the granular flow, while the overall distribution of power can be affected by the details of the simulation.

Thus, regardless of our choice of 3D hydrodynamical code, model~I is the preferred background model, and the free exponent is unchanged (within the uncertainties). However, we see differences in the details of the fit. In order to investigate if these differences between the Reference \cob\ power spectrum and the \sta\ power spectrum can be reproduced by changing key characteristics of the simulation, we study the effects of changing the spatial resolution and the total duration of the simulation in the following.

\subsection{Effect of spatial resolution}
\label{subsec:resolution}

The second test, detailed in this subsection, concerns the potential effect of the spatial resolution of the simulation, in particular on the parameters of background model I. In order to investigate this, we use models d, g, j, and c600 (see Table~\ref{tab:simulations}) in addition to our Reference model. Note that to distinguish between the 3D hydro models and the employed background models (in Table~\ref{tab:models}), we use lower-case letters for the hydro models (such as d and j) and upper-case letters (such as H and I) for the background models.

Models d, g and j are identical in set-up and d and g comparable in duration, but the spatial resolution is different. Model d has a lower spatial resolution (the size of the box with respect to the number of grid cells along a given dimension) than model g that again has a lower spatial resolution than model j. Model c600 has a short duration and small simulation box. The spatial resolution is comparable to model j, but the treatment of radiative transfer is more detailed than in the other \cob\ models.

Returning to Fig.~\ref{fig:comparingpower}, it can be seen that the slope at high frequencies (around $5-6$ mHz) is similar in the different power spectra, but that the 'flattening' at the highest frequencies occur earlier the lower the spatial resolution of the simulation is. However, the most noticeable difference between the power spectra is the amount of high frequency power (note that the power spectra have been scaled to match in the $1-2$ mHz region). We see that the power spectra from model d have the least power around $5-6$ mHz, followed by the Reference model power spectrum, the power spectrum from model g, the power spectrum from model j and then the \sta\ and c600 ones. This is consistent with an overall trend that the amount of high frequency power increases with increasing horizontal spatial resolution, that is the better the surface granules are resolved in the simulation. An exception to this relation is the \sta\ and c600 simulations, which are close in both power around $5-6$ mHz and horizontal spatial resolution with \sta\ having the highest resolution and c600 the largest amount of power. However, given how close they lie and the differences between the two simulations, this is not a cause for concern.

We have fitted background model I to each of the four power spectra: model d (integrated), model g, model j and model c600, where we cut the power spectra at the Nyquist frequency of the Reference power spectrum in order to be able to compare the fits directly. However, the flattening of the model d power spectrum already becomes pronounced at ${\sim}6$ mHz because of the low spatial resolution. As a consequence, we decided to cut this spectrum at $8$ mHz instead. For power spectra from hydro models d and g we made fits including various numbers of Lorentzians to account for the box-modes and picked the fit with the highest marginal likelihood. The values of the characteristic frequencies and the free exponent can be seen in the bottom two lines of Table~\ref{tab:fitparams}.

In the case of model c600, we chose to not bin the power spectrum because of the low number of points. However, this results in a fairly noisy power spectrum, which shows no evidence of two characteristic frequencies. Thus, the result from fitting background model I to the c600 power spectrum is not very convincing, but it is worth noting that the slope at high frequencies (the free exponent) is consistent with a value around six ($k=6.04 \pm 0.09$). The same is true in the case of model j ($k=5.92 \pm 0.08$).

From Table~\ref{tab:fitparams} it is evident that the characteristic frequencies are sensitive to the differences between the hydro models with a trend that d is increasing with increasing spatial resolution. This is also in agreement with what we see from the fit to the \sta\ model. Furthermore, the free exponent in the background description appears to be fairly robust against changes in the spatial resolution. Even in the case of hydro model d, which has the lowest spatial resolution, the free exponent is less than $2 \sigma$ smaller than for the Reference model and \sta. Thus, we can conclude that the differences between our Reference power spectrum and the one from \sta\ is likely due to a different spatial resolution. Furthermore, it seems that unless the spatial resolution is very low, it will not have an impact on the high frequency slope in the power spectrum.

\subsection{Effect of changing the duration}
\label{subsec:duration}

\begin{figure}
	\includegraphics[width=\columnwidth]{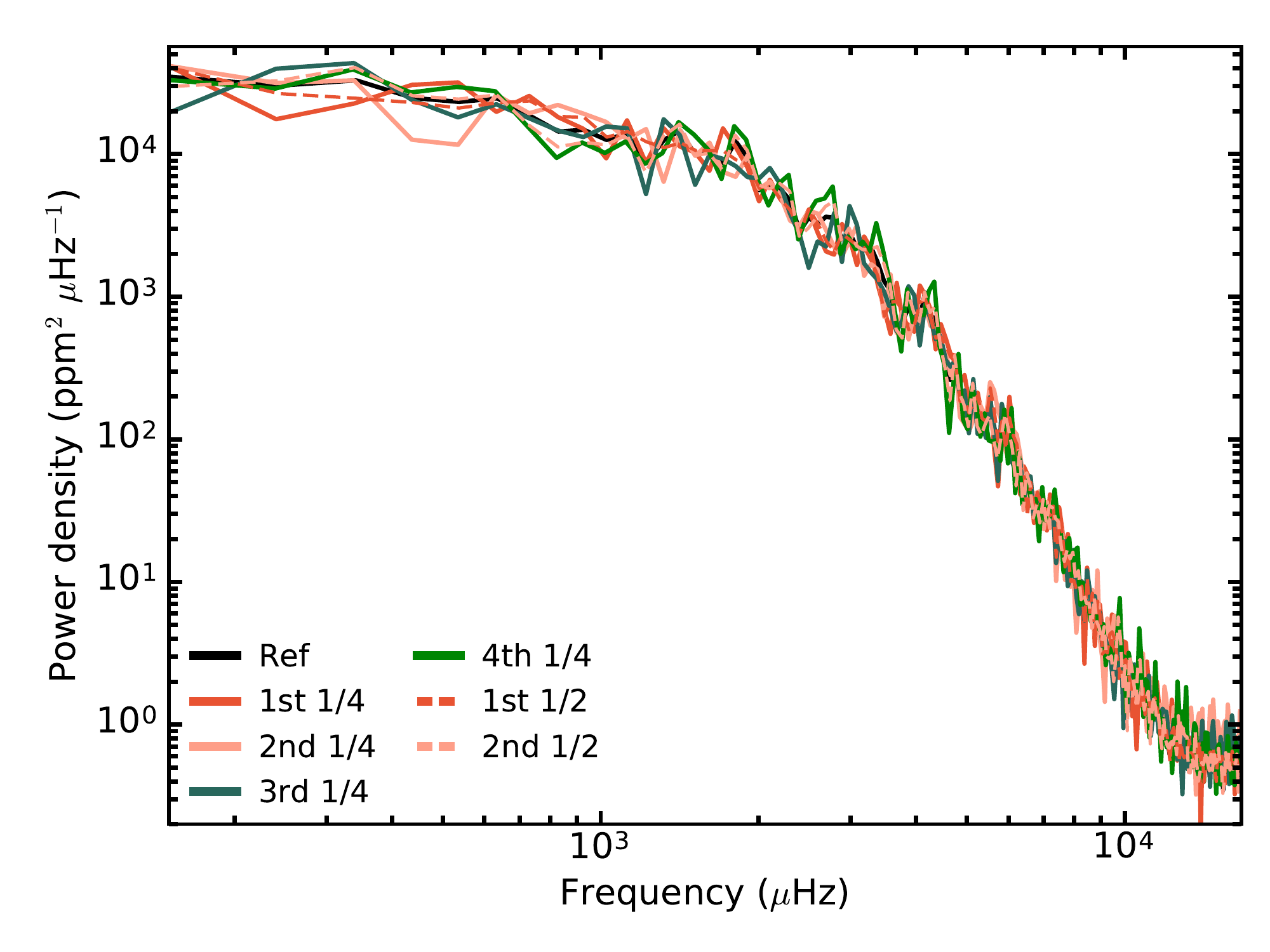}
    \caption{Smoothed version of power spectra computed from segments of the Reference \cob\ time series. The time series have lengths of 25 percent and 50 percent of the total duration. It can be seen that the duration and the actual segment of the time series used to compute the power spectrum only has a very minor impact on the shape of the spectrum.}
    \label{fig:durationpower}
\end{figure}

The third test that can be done is to study the effect of changing the length of the time series used to compute the power spectra. We split the time series of the Reference \cob\ simulation into shorter segments (eights, quarters, thirds and halves) based on the total duration and computed power spectra for each of these segments. It should be noted that since the last ${\sim}3$ percent of the Reference time series is computed with a higher sampling rate, the end segments will contain more points than the other ones. 
The power spectra for the four quarters, each with a duration comparable to that of the \sta\ simulation, and the two halves are plotted in Fig.~\ref{fig:durationpower}, where it can be seen that the overall appearance of the power spectrum is only affected to a small extent by changing the duration of the underlying time series. In particular, it can be noted that the differences between the power spectra in Fig.~\ref{fig:comparingpower}, most importantly the amount of high-frequency power, cannot be attributed to the varying length of the time series used.

Background model I was fit to each of the 17 power spectra in order to investigate how changing the duration impacts the parameter estimates. From this it is evident that the free exponent $k$ and to a lesser extent the characteristic frequency $d$ show an overall positive correlation with the length of the segment. However, $d$ and $k$ are positively correlated (see Fig.~\ref{fig:modI_corner}), so it is not surprising that both parameters show this trend. When studying the individual fits to the power spectra, it is evident that lower power levels at very high frequencies (beyond ${\sim}10$ mHz) result in higher values of $k$. This makes sense because a steeper slope can be 'accommodated' by the fit, when the amount of power at the highest frequencies is lower.

\begin{figure}
	\includegraphics[width=\columnwidth]{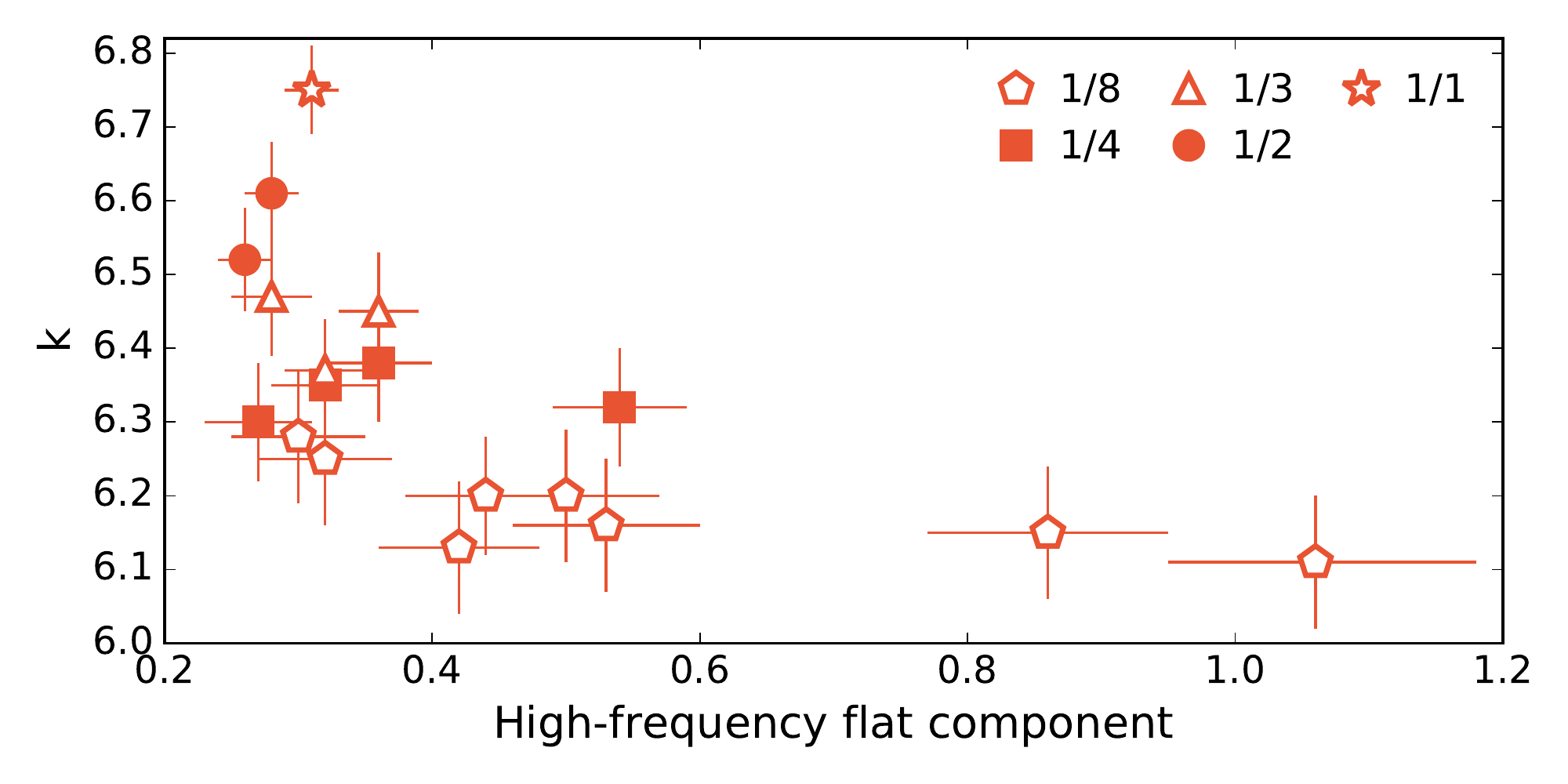}
    \caption{Value of the free exponent $k$ as a function of the value of the flat component that was added to background model I. The different symbols indicate the duration of the segment used to compute the power spectrum (see the legend for details). The anti-correlation is evident along with a trend that the shorter the time series segment is, the lower is the resulting value of the fitted free exponent $k$.}
    \label{fig:powerk}
\end{figure}

\begin{figure*}
	\includegraphics[width=.7\textwidth]{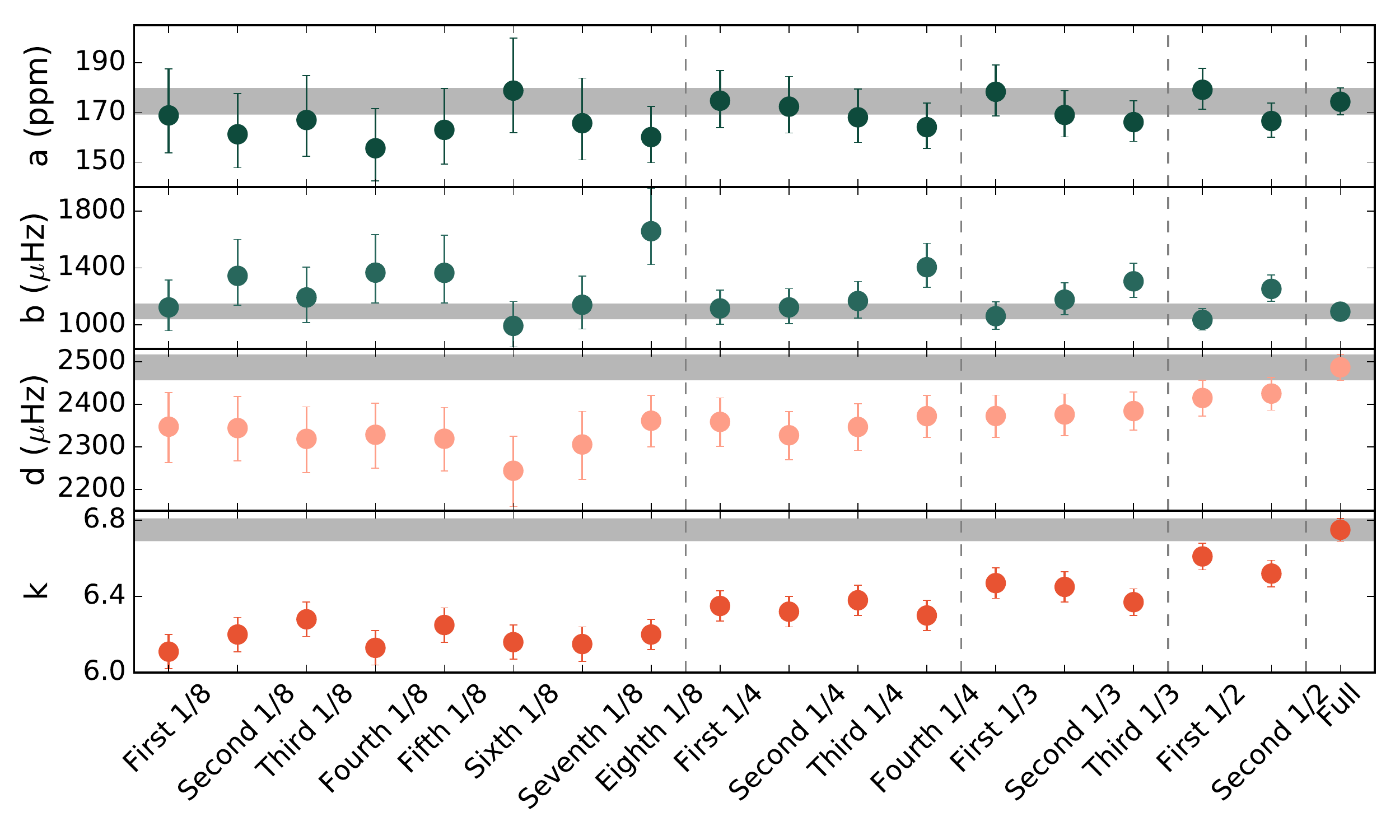}
    \caption{Value of the amplitude (top), characteristic frequencies (middle) and free exponent (bottom) when adding a flat component to background model I. The plot shows the derived values for each power spectrum indicating which segment of the time series that was used to compute it. The shaded area in each panel gives the value (including $1\sigma$ uncertainties) found for the full time series. The vertical dashed lines separate results for different segment lengths.}
    \label{fig:duration}
\end{figure*}

In order to confirm this degeneracy, we added a flat (i.e. frequency independent) component to model I to capture the amount of power at very high frequencies. This led to fits that have a higher marginal likelihood and display the anti-correlation between the power and the exponent $k$, as depicted in Fig.~\ref{fig:powerk}. The plot further hints at the above-mentioned correlation between the segment length and the value of $k$, independent of the amount of highest-frequency power in the power spectrum. This, along with the impact on the other model parameters of changing the duration of the time series, is also visible in Fig.~\ref{fig:duration}, which shows the parameter estimates and uncertainties for each of the computed power spectra.

It is clear from the figure that the values of the amplitude and the first characteristic frequencies are consistent regardless of which segment is used, with the values from the shorter segments showing larger uncertainties as expected. While the second characteristic frequency and the free exponent also show larger uncertainties for the shorter segments, they also exhibit a step-like behaviour or an overall degeneracy between segment length and the parameter value as noted above. This could hint at a missing high-frequency component in the model I background model. However, as power spectra of real stars display white noise in this part of the spectrum, any component at these high frequencies will be negligible compared to the white noise component. We will address the effect of white noise in the following.

\subsection{Adding noise}
\label{subsec:noise}

\begin{figure}
	\includegraphics[width=\columnwidth]{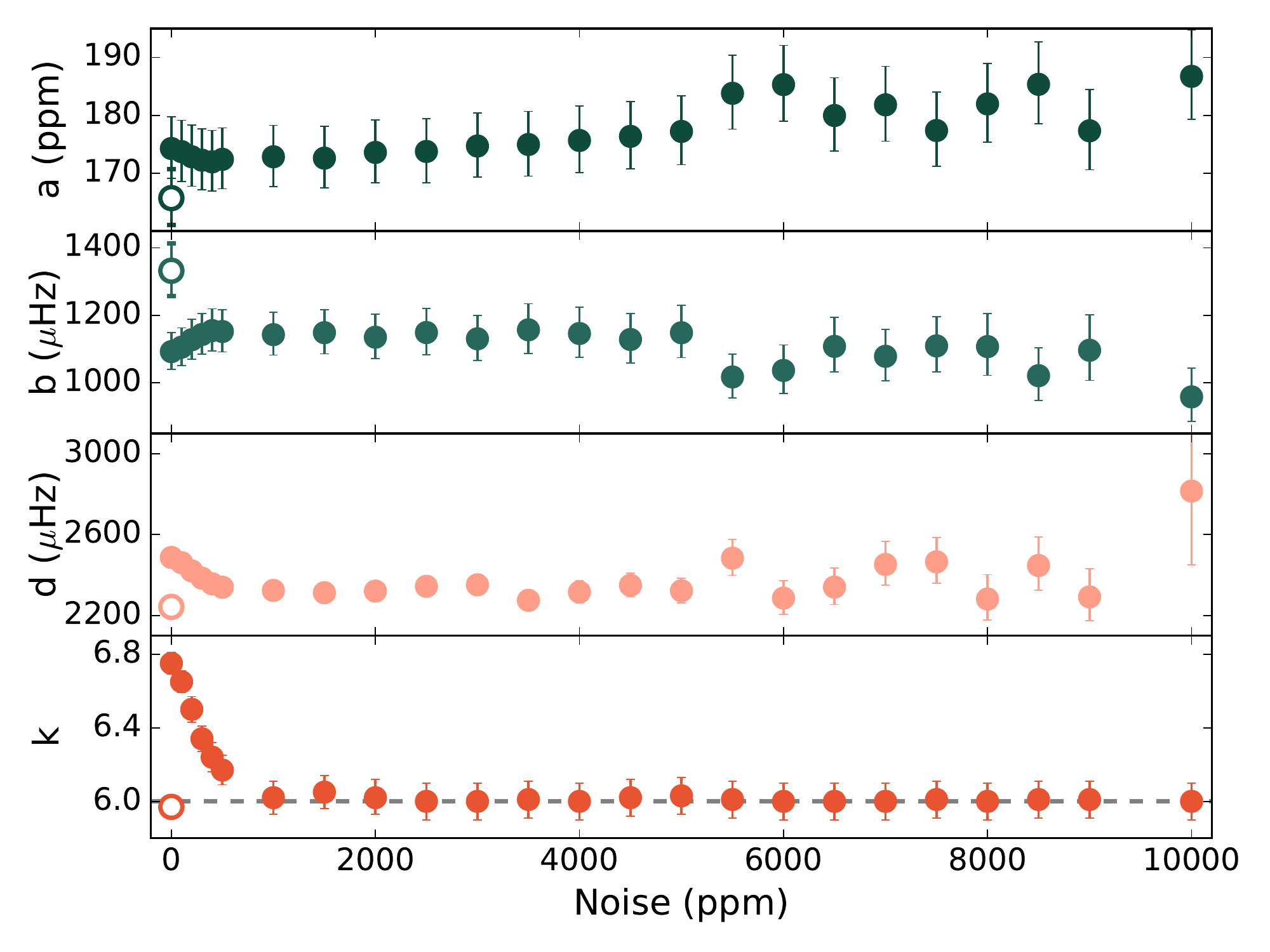}
    \caption{Value of the amplitude, characteristic frequencies and free exponent of background model I with an added white-noise component as a function of the amount of white noise added to the \cob\ time series. Top panel shows the amplitude, middle panels show the frequencies and the bottom panel gives the free exponent. The open circle in each panel shows the Model I results without any noise for reference (from Sect.~\ref{subsec:cobresults}). Also shown is the noise-free Reference time series fitted with a model including a white-noise component (same results as those shown as the outermost point in Fig.~\ref{fig:duration}). The dashed line in the bottom panel shows an exponent of $6.0$.}
    \label{fig:effectnoise}
\end{figure}
The last test to address is the use of background model I and the Reference \cob\ time series to investigate the effect that adding white noise to the time series has on the model parameters. This is interesting since a real star would show this component in the power spectrum. We used the same variant of model I as introduced in Sect.~\ref{subsec:duration}, namely model I plus a flat component. The added noise was Gaussian white noise with zero mean, which was added to the relative flux time series (having fluctuations on a scale of $6.5 \cdot 10^3$ ppm). The standard deviation of the normal distribution, that the noise was drawn from, was varied between $0$ and $10^4$ ppm. The impact on the determined parameters of model I can be seen in Fig.~\ref{fig:effectnoise}.

Overall, the effect of adding noise is that the scatter and uncertainties increase, as is evident from the top three panels in the figure. Of these, the characteristic frequency $d$ is the one that shows the largest scatter. This can be attributed to the fact that as the noise level increases, it is the second 'knee' on the curve that is most prominently washed away, and thus $d$ becomes much harder to determine. Further, the preceding section showed that $d$ exhibits some degree of correlation with the free exponent $k$, something which can also be seen in Fig.~\ref{fig:effectnoise} at the lowest noise levels. In fact, the behaviour at the lowest noise levels is similar to what was found in Sect.~\ref{subsec:duration}, where $d$ and more prominently $k$ show the trend of lower values as the value of the flat component of model I (the white noise) increases. That this trend with $k$ is similar, can be confirmed by comparing the left side of the bottom panel in Fig.~\ref{fig:effectnoise} to Fig.~\ref{fig:powerk}. Thus, when the noise level is negligible compared to the intrinsic scatter (below ${\sim} 1000$ ppm), we see the degeneracy between $k$ and the flat component of the background description (added to account for the noise) that was also evident in Fig.~\ref{fig:powerk}.

As the noise level increases, the value of $k$ settles around six, which is fully consistent with the standard model I no-noise case and the other simulations described in Sects.~\ref{subsec:stagger} and \ref{subsec:resolution}. This supports the above statement, that once the white noise dominates over this 'extra' component mentioned in the previous section (from ${\sim}1000$ ppm), i.e. when the simulated power spectrum becomes more 'real star-like', then $k$ is essentially constant.

Thus we can conclude that adding white noise, even at levels comparable to and greater than the intrinsic scatter in the time series, does not have a significant impact on the parameter values for background model I. It is not the goal of this analysis to determine the free parameter $k$ as it will be estimated when using background model I, but the results suggest that for real stars with realistic noise levels, it will be in the vicinity of $6$.

\section{Discussion}
\label{sec:discussion}

Based on background models A\&E, B\&F and C\&G presented in Sect.~\ref{subsec:cobresults}, a two component model is not preferred over a one component model, something which is not the case for real stars observed by \textit{Kepler} \citep{ref:karoff2013,ref:kallinger2014}. That our simulated data lend little support to a two-component model can be seen by eye upon inspection of Fig.~\ref{fig:modelfit}, since the Reference power spectrum displays no kink at ${\sim}1500 \ \muup$Hz (half the solar $\nu_\mathrm{max}$), a depression known to arise if the background in this region of the power spectrum is composed of two components \citep{ref:kallinger2014}. In fact, inferring from the model probabilities in Table~\ref{tab:results}, the only model that provides a satisfactory match to the simulated power spectrum is model I, a model which was not included in either of the above-mentioned works.

In order to investigate how our findings hold up in a real star and to compare our results to a single star from \citet{ref:kallinger2014}, we used our set-up to test the different background models listed in Table~\ref{tab:models} on the red-giant \textit{Kepler} star KIC~7949599. Here, we included in the total model the background model (without mirroring around the Nyquist frequency), an activity component (similar to model A), a Gaussian envelope (for the p-modes) and a flat white-noise contribution in addition to the apodization factor. We find that background model I is still the preferred background description when considering the marginal likelihood (model evidence) for each model, in agreement with the results obtained from the simulations.

For the Reference \cob\ power spectrum, the parameter estimate of the free exponent ($k$) in background model I is $5.97 \pm 0.03$ (refer to Table~\ref{tab:fitparams}). This result is not far from the values found by \citet{ref:harvey1993} when studying Ca II K line observations of the solar chromosphere (5.0 and 5.6 associated with chromospheric bright points and granulation overshoot respectively). Additionally, \citet{ref:karoff2012} studied the solar power spectrum up to 3200 $\mu$Hz (from Virgo@SOHO). Here, he found an exponent of 6.2 for the faculae component, something which is also in agreement with our determined value.

It is interesting to note that the second of the free exponents (the one relevant at the higher frequencies, as the free exponent in model I) in model H turns out to be in the vicinity of six ($5.75 \substack{+0.04 \\ -0.03} $) when fitting this model to the Reference \cob\ power spectrum. This is similar to the free exponent of model I, but it is in contrast to an exponent close to four, which was found by \citet{ref:kallinger2014} in their fits to red giants from \textit{Kepler}. However, when we fit the star KIC~7949599, which is the one detailed in the paper by \citet{ref:kallinger2014}, our results for model H agree with theirs within $1\sigma$ for most parameters (and within $2\sigma$ for the rest), while model I still retains an exponent significantly closer to six. It is unclear whether this difference is caused by the different evolutionary state of the star (the Sun as compared to a red giant), to a difference between the simulation and the real data or something else. 

In regards to the duration of the time series, it is worth noting that the value of the free exponent determined using the \sta\ simulation is in agreement with that found from the Reference \cob\ one, although the duration of the former is similar to the 'quarter-length' time series discussed in Sect.~\ref{subsec:duration}. This can be understood by considering the spatial resolution. 
We have seen in Sect.~\ref{subsec:duration} that when the time series segments are shorter, the amount of power at the highest frequencies increase along with a decrease of $k$. However, in the case of the \sta\ model the higher spatial resolution in comparison to the \cob\ model means that the flattening at the highest frequencies occur at even higher frequencies than for the \cob\ model. This in turn likely means that small shifts in the overall power level at the highest frequencies, do not have as large an effect on the fitting.

\section{Conclusion}
\label{sec:conclusion}

In this paper we have investigated the granulation background in a \cob\ simulation of the Sun. We have found that the background is best described by an expression containing a single power level and two characteristic frequencies: $\frac{a}{1 + (\nu/b)^2 + (\nu/d)^k}$ (our model I) with the free exponent $k$ taking on a value around six.

We have investigated how this conclusion is affected by using different simulations, having different spatial resolutions or different durations, and the conclusion is that the result is robust to these effects. Additionally, we added noise to our \cob\ simulation, which except for the cases with very low noise, also does not change our finding. Thereby, we suggest that model I, at least for a simulated Sun, gives a good representation of the granulation background seen in power spectra.

However, the difference between our simulated power spectrum fitted with model I and the well-established results found for real stars by, for instance, \citet{ref:kallinger2014} and \citet{ref:karoff2013} is striking and something that warrants further investigation. It could simply be because we only studied a single simulated star, where the other works dealt with more stars, but the simulation and real stars are also rather different, for instance in terms of magnetic fields, activity and stellar oscillations. The difference could also point to the fact that the simulations are not perfect and maybe do not reproduce the granulation background in the power spectrum as well as expected. This could be studied further by carrying out a similar analysis using simulations of, for instance, a red-giant star, and by comparing model I and model H for a suite of real stars.

\section*{Acknowledgements}

The authors wish to thank Guy R. Davies for valuable discussions and python routines implementing PyStan, Ren{\'e} Salhab for sharing python routines for reading \cob\ files, and Matthias Steffen for providing the solar model c600.
Additionally, the authors would like to thank the anonymous referee for providing comments that improved the manuscript. M.S.L. is supported by the Carlsberg Foundation (Grant agreement no.: CF17-0760). Funding for the Stellar Astrophysics Centre is provided by The Danish National Research Foundation (Grant DNRF106). H.G.L. acknowledges financial support by the Deutsche Forschungsgemeinschaft (DFG, German Research Foundation) -- Project-ID 138713538 -- SFB 881 (``The Milky Way System'', subproject A04). The simulations of our Reference model were carried out at CINECA (Bologna/Italy) with CPU time assigned under INAF/CINECA agreement 2008/2010.

\section*{Data Availability}

The data underlying this article will be shared on reasonable request to the corresponding author.




\bibliographystyle{mnras}
\bibliography{simsun_gran} 



\appendix

\section{Difference in shape between patch and disk-integrated power spectrum}
\label{app:shape}

\begin{figure}
	\includegraphics[width=\columnwidth]{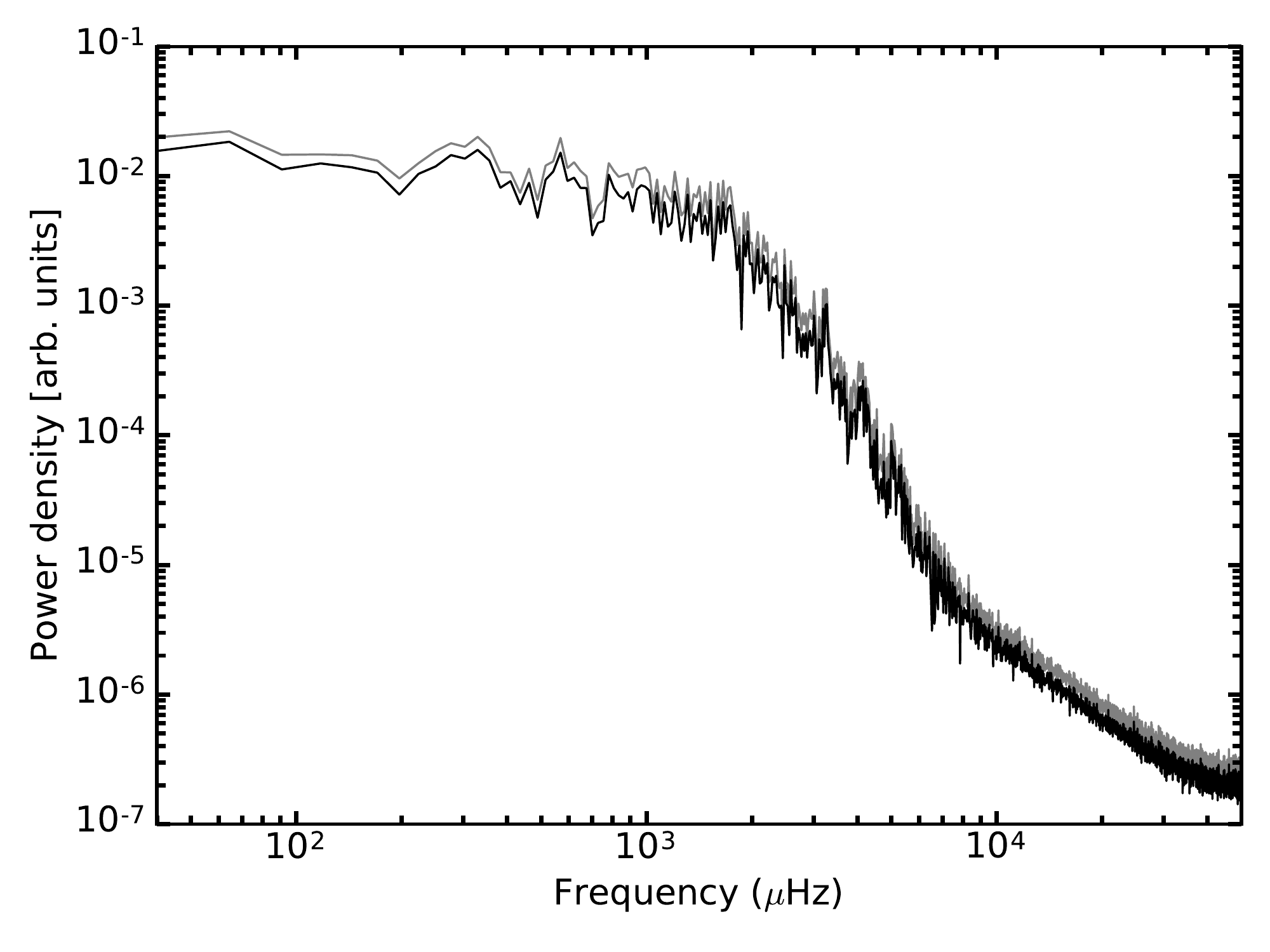}
    \caption{Smoothed simulated power density spectrum of hydro model d computed using a single patch (black) and disk-integrated in a manner following \citet{ref:ludwig2006} (grey). It can be seen that the overall shape is unchanged.}
    \label{fig:app:shape}
\end{figure}
In this work, we assume that using the power spectrum computed from just a single patch as opposed to a disk-integrated one will not affect our results. In order to test this, we compared the result of fitting the model I background to two different power spectra both coming from the same three-day \cob\ solar simulation, but one based on the average horizontal flux in the box (\textit{patch}) and one modified following \citet{ref:ludwig2006} (\textit{disk-integrated}). The two power spectra can be seen in Fig.~\ref{fig:app:shape}, where a small overall shift of the power level can be seen \citep[as discussed in][]{ref:ludwig2006}.

In comparison to the Reference \cob\ power spectrum analysed in the main text, the power spectra in Fig.~\ref{fig:app:shape} show more prominent box modes. In order to account for this, we fitted the model I background and a Lorentzian profile simultaneously to allow for the fitting of the dominant box mode at $\sim{3.1}$~mHz.

All of the fitting parameters are found to be consistent within $1\sigma$, in particular the free exponent is $5.8$ in both cases, which is in excellent agreement with what was found from the other solar models examined in this work. Thus, we can confirm our assumption that the overall shape of the granulation background, described by the free exponent and the characteristic frequencies, is not affected to a detectable amount by our choice to work with the power spectrum from a single patch rather than scaling it to a full disk.


\section{Priors used in the Bayesian approach}
\label{app:priors}

As stated in Sect.~\ref{subsec:fitting}, we employed weakly informative log-normal priors for the amplitudes and characteristic frequencies and normal distributions for the free exponents. In the case of the Nyquist frequency, we employed normal prior centred on the Nyquist frequency estimated from the simulation with a width of $0.001$ times the estimated frequency because of the very well-defined value.

Table~\ref{apptab:priors} gives the priors used in the model comparison for the Reference simulation.

\begin{table}
	\centering
	\caption{Priors used for the Reference \cob\ simulation.}
	\label{apptab:priors}
	\begin{tabular}{lccc} 
		\hline
		Parameter & Distribution & Central value & Width\\
		\hline
		a	 		 	 & log-normal & $ 1.52 $ & $ 0.3 $ \\
		b				 & log-normal & $ 2.58 $ & $ 0.3 $ \\
		c				 & log-normal & $ 1.53 $ & $ 0.3 $ \\
		d				 & log-normal & $ 3.38 $ & $ 0.3 $ \\
		l				 & normal & $ 4.0 $ & $ 0.1 $ \\
		k (Model H)		 & normal & $ 4.0 $ & $ 0.1 $ \\
		k (Model I)		 & normal & $ 6.0 $ & $ 0.1$ \\
		$\nu_\mathrm{Nyq}$ & normal & $ 1.665 \cdot 10^4 $ & $ 0.001 $ \\
		\hline
	\end{tabular}
\end{table}

\bsp	
\label{lastpage}
\end{document}